\documentclass[twocolumn,amsmath,amssymb,superscriptaddress]{revtex4}
\usepackage{graphics}
\usepackage{fancybox}
\usepackage{color}
\usepackage{physics}
\usepackage{amsmath}
\usepackage{enumerate}
\usepackage[normalem]{ulem}

\usepackage{amsmath, amsthm, amssymb}

\usepackage[utf8]{inputenc}
\usepackage[english]{babel}

\newcommand{\dr}[1]{\dd{\vec{#1}}}
\newcommand{\rvec}[1]{(\vec{#1})}

\newcommand{\cc}{\text{c.c.}}
\newcommand{\create}{\hat{c}^{\dagger}}
\newcommand{\destroy}{\hat{c}}
\newcommand{\vmel}[2]{\big(#1\big|#2\big)}
\newcommand{\mell}[2]{\langle #1 || #2 \rangle}
\renewcommand{\r}{(\vec{r})}
\renewcommand{\vec}{\mathbf}

\begin{document}

  \cornersize{.3}

  \newtheorem*{corollary}{Corollary}
  \newtheorem*{lemma}{Lemma}
  \newtheorem*{remark}{Remark}

 \title{Optimal power series expansions of the Kohn-Sham potential}

 \author{Timothy J. Callow}
 \affiliation{Department of Physics, Durham University, South Road, Durham, DH1 3LE, United Kingdom}
 \affiliation{Max-Planck-Institut f\"{u}r Mikrostrukturphysik, Weinberg 2, D-06120 Halle, Germany}
 \author{Nikitas I. Gidopoulos}
 \affiliation{Department of Physics, Durham University, South Road, Durham, DH1 3LE, United Kingdom}

  \date{\today}
  \begin{abstract}
    A fundamental weakness of density functional theory (DFT) is the difficulty in making systematic improvements to
    approximations for the exchange and correlation functionals. In this paper, we follow a wave-function-based approach
    [N.I. Gidopoulos, Phys.~Rev.~A,~{\bf 83},~040502~(2011)] to develop perturbative expansions of the Kohn-Sham (KS)
    potential. Our method is not impeded by the problem of variational collapse of the second-order correlation energy
    functional. \\
    Arguing physically that a small magnitude of the correlation energy implies weak perturbation and hence fast convergence of the perturbative
    expansion for the interacting state and for the KS potential, we discuss several choices for the zeroth-order Hamiltonian in such expansions.
    Our first two choices yield KS potentials containing only Hartree and exchange terms: the exchange-only optimized effective
    potential (xOEP), also known as the exact-exchange potential (EXX), and the Local Fock exchange (LFX) potential.
    Finally, we choose the zeroth order Hamiltonian that corresponds to minimum magnitude of the second order correlation energy,
    aiming to obtain at first order the most accurate approximation for the KS potential
    with Hartree, exchange and correlation character.
    \end{abstract}
    %
  %
  \maketitle
  \section{Introduction}

  Electronic structure calculations are becoming indispensable in many areas of modern science, with applications spanning fields from drug discovery~\cite{Drugs} to superconductivity~\cite{superconductivity}. This change has been largely driven by the continuing development of density functional theory (DFT) over 50 years, which enjoys extraordinary and growing popularity~\cite{Burke_review}; and by the robustness of modern computational codes combined with the increasing speed of modern computers.

  The importance of DFT in the theory of electronic structure was reflected in the 1998 divided Nobel
  Prize in Chemistry between W. Kohn~\cite{Kohn_Nobel_lecture} and J. Pople~\cite{Pople_Nobel_lecture} for their developments of DFT and of computational
  methods in quantum chemistry respectively. The shared prize also reflected the importance of the
  synthesis of the two theories, since on their own both are limited, either by the difficulty to improve
  systematically on the approximations (DFT), or by poor scaling (wave function theory -- WFT).
  In order to overcome these limitations, and to satisfy the growing demands for more accurate electronic structure calculations, on larger and more complicated systems, it is
  important to gain new insights.
  Such
  insights can be obtained from the integration of
  DFT with
  WFT~\cite{engel1999explicit,Engel_vdw,Grabowski2,abinitioDFT,Sanchez_Wu,Schweigert_ab_init_correlation_funcs,grabowski,Bartlett}.

  In the scientific community, a dichotomy is perceived between DFT and WFT.
  The different emphasis of the two theories, density vs wave function, appears
  to hinder their smooth integration.
  The constrained search formulation by Levy~\cite{Levy_constrained} and by Lieb~\cite{Lieb_constrained} and the adiabatic connection path
  construction~\cite{ad_conn_1,ad_conn_2,ad_conn_3} are seminal works in this area.
  Put together, they enabled G{\"o}rling and Levy
  to formulate DFT perturbation theory (PT)~\cite{GL_PT_1,GL_PT_2}, and Bartlett and co-workers to develop ab initio DFT~\cite{Grabowski2}.
  In these approaches, the correlation energy is approximated from second-order
  PT (or higher) and the KS potential is then determined using the optimized
  effective potential (OEP) method~\cite{OEP1,OEP2}.

  However, as the correlation energy from second-order PT is unbound from below,
  any minimization of the subsequent total energy functional is variationally
  unstable, tending to yield unphysically low total energies~\cite{Sanchez_Wu,rohr2006variational}.
  Of course, there are physical situations (such as molecular dissociation) where
  the second order correlation energy term from a single reference Slater
  determinant will necessarily diverge to negative infinity;
  but the point is that using a correlation energy functional from second order
  PT, the tendency to diverge is inherent for all systems.
  In practice this divergence indeed turns out to be far more common than in
  quantum chemical methods employing perturbation theory, such as second order
  M{\o}ller-Plesset PT (MP2)~\cite{bartlett_collapse,rohr2006variational}.
  Whilst ways to alleviate the variational collapse have been put forward,
  for example by using Fock exchange energies instead of the true KS orbital energies~\cite{abinitioDFT,Schweigert_ab_init_correlation_funcs},  
  the absence of a rigorous solution to this issue has hindered
  progress in the cross-fertilization between WFT and DFT.
 Nevertheless, many-body perturbation theory (MBPT) has been employed successfully to yield an accurate 
 correlation energy functional for DFT in the random phase approximation (RPA) 
 \cite{LANGRETH19751425,Perdew_RPA,Bartlett_RPA,Gross_RPA,Gorling_RPA_2012,Gorling_RPA2,nguyen_2009,RPA_review,rinke_2012}, 
 by combining the adiabatic connection path construction with the fluctuation dissipation theorem \cite{kubo}, 
 but also using the Sham-Schl\"uter equation \cite{godby_sham_schluter,sham_schluter}.
  
  A few years ago, Gidopoulos proposed a natural way to integrate DFT and WFT, by constructing a
  pure WFT method whose solution happens to be the Kohn-Sham (KS) system of DFT~\cite{gidopoulos1}.
  In that theory, it is no longer necessary to fix the electron density along the adiabatic connection path or elsewhere,
  making it a straightforward task to employ techniques from WFT in order to determine key quantities
  of DFT.
  The aim of the present paper is to demonstrate how the new formalism works, by
  constructing perturbative expansions of the KS potential that can be expected
  to converge optimally.

  The paper is structured as follows. In section 2, we review the WFT method from~\cite{gidopoulos1}, which is based on the minimization of the energy difference given by,
  \begin{equation} \label{eq1}
    T_\Psi [v] \doteq
    \langle \Psi | H_v | \Psi \rangle - E_v > 0,
  \end{equation}
  where $\Psi$ is the ground-state (g.s.) of the physical (interacting) system, and $H_v$ is an effective Hamiltonian,
  \begin{equation} \label{eq2}
    H_v = \sum_{i=1}^N
    \left[ - \frac{1}{2} \nabla_i^2  + v_{en} ({\bf r}_i ) + v({\bf r}_i ) \right],
  \end{equation}
  for some local potential $v\r$, which simulates the electron-electron
  repulsion.
  The g.s. of $H_v$ is $\Phi_v$ and the g.s. energy is $E_v$,
  \begin{equation} \label{eq3}
    H_v \Phi_v = E_v  \Phi_v.
  \end{equation}

  The energy difference $T_\Psi [v]$ is strictly positive due to the Rayleigh-
  Ritz inequality; the positivity of the energy difference is preserved
  even when it is expanded with PT and an approximation up to  second
  order is kept. Hence when $T_\Psi [v]$ is minimized there is no
  possibility of incurring the variational collapse of DFT with a correlation
  energy functional from second order PT.
  %
  %
  The relation between Eq.~\eqref{eq1} and Lieb's functional \cite{lieb_1983} is
  explored in Ref.~\cite{teale_2017}.

  In section 3, we see how the the optimization over the total energy in the traditional OEP manner 
  is equivalent to optimizing over the magnitude of the correlation energy.
  We then compare in section 4 our method with the traditional DFT perturbation theory (DFT PT)
  approach. In section 5, we discuss three different expansions for the KS
  potential: the first two yield at first order the exact exchange and local Fock
  exchange potentials respectively, and the final one already at first order
  includes correlation and has not been considered in the literature so far.
  Finally, we draw conclusions in section 6.

  \section{Power series expansions of the KS potential}

  In this section, we review the key results from the WFT approach developed by
  Gidopoulos in~\cite{gidopoulos1}; namely, how minimization of the energy
  difference in Eq.~\eqref{eq1} yields the KS potential, and how to derive power
  series expansions of the KS potential from perturbation theory.

  Ineq.~\eqref{eq1} holds because the interacting state $\Psi$ cannot be the exact
  g.s. of a non-interacting Hamiltonian $H_v$; however, we can view $\Psi$ as an
  approximate g.s. of $H_v$. Then, choosing $v\r$ to minimize $T_\Psi[v]$ amounts
  to selecting the Hamiltonian $H_v$ in the class~\eqref{eq2} which optimally
  adopts $\Psi$ as its approximate ground state. 
  It transpires that the minimizing 
  potential $v_s$ of $T_\Psi[v]$ is the KS potential, since 
  setting the functional derivative of $T_\Psi [ v ] $ w.r.t. $v \r$ equal to zero yields
  \begin{equation}
      \rho_\Psi \r - \rho_{s}\r=0,
  \end{equation}
  where $\rho_\Psi$ is the density of $\Psi$ and $\rho_s$ is the density of $v_s$. 
  By the definition of the KS potential and the Hohenberg-Kohn theorem, the potential $v_s$ must be the KS potential 
  (a detailed proof can be found in \cite{gidopoulos1}).
\color{black}

  With the variational principle (1), the problem of constructing a power series
  expansion of the KS potential is simplified, as it is
  no longer necessary to employ the adiabatic connection path formalism,
  where the local potential varies in an unknown manner along the path.
  Instead, we may substitute any power series expansion of $\Psi$ in $T_\Psi [v]$, truncating the
  energy difference $T_\Psi [v]$ at a finite order. Optimization over $v$ for a given expansion of $T_\Psi[v]$ 
  then yields a corresponding expansion for the KS potential.

  Of course, for a specific power series expansion of $\Psi$, it was always possible to truncate the expansion at any order and thus 
  obtain its density; numerically inverting the density then leads to a (numerical) power series expansion of the KS potential.
  The difference with the present theory is that this procedure can be formally carried out for a whole class of
  Taylor series expansions of $\Psi$, characterized by the choice of zeroth-order
  Hamiltonian. It is then possible to consider the corresponding class of Taylor series expansions of the KS potential and
  search in that class for those expansions that converge faster than others. In
  other words, our method allows us to construct and then search a wide space of
  power-series expansions for the KS potential, to find those expansions which
  are expected to be the most accurate when truncated at some finite order.

  In the following, we review from~\cite{gidopoulos1} the way to construct the lowest order in such expansions.
  In order to expand the energy difference, we use the interacting state $\Psi_u (\alpha)$, g.s.  of
  the perturbative Hamiltonian $H_u (\alpha)$:
  \begin{eqnarray}
    &&   H_u (\alpha) \, \Psi_u (\alpha)  =  E_u (\alpha) \, \Psi_u (\alpha) ,  \label{3}
    \\
    && H_u (\alpha) = H_u + \alpha \, \big[ V_{ee} - \sum_i u ({\bf r}_i) \big]  . \label{4}
  \end{eqnarray}
  The zeroth-order Hamiltonian is $H_u$; it belongs to the class of effective
  Hamiltonians (\ref{eq2}) but with a local potential
  $v_{en} ({\bf r}) + u({\bf r})$ instead of $v_{en} ({\bf r}) + v({\bf r})$.
  Similarly to $v({\bf r})$, the effective potential $u({\bf r})$ mimics the electronic repulsion in a mean-field way.
  The fully interacting Hamiltonian $H$ is obtained for $\alpha = 1$, $H_u (1) = H$.

  Obviously, for $\alpha = 0$, $\Psi_u (0) = \Phi_u$.
  If we substitute $\Psi_u (0)$ in place of $\Psi$ in $T_\Psi [v]$ and search
  for the potential that minimizes $T_{\Psi_u (0) } [v]$, the minimizing
  potential will be $v=u$ obviously.
  Hence, for small $\alpha$, we expect that the potential which minimizes $T_{\Psi_u (\alpha) } [v]$ will be close to $u$.
  Setting
  \begin{equation} \label{eq6}
    v ({\bf r}) = u ({\bf r})  + \alpha v' ({\bf r}) ,
  \end{equation}
  the leading term in the energy difference
  $T_{\Psi_u (\alpha)} [ u + \alpha v ' ]$
  turns out to be of second order:
  \begin{equation} \label{eq7}
    T_{\Psi_u (\alpha)} [ u + \alpha v ' ] = \alpha^2 T_u [ u + v'] +
    \mathcal{O} (\alpha^3)  ,
  \end{equation}
  where
  \begin{equation} \label{eq8}
    T_u [ w] = \sum_{n \ne {\rm g.s.}} { | \langle \Phi_{u , n } | V_{ee} - \sum_i w ({\bf r}_i ) | \Phi_u \rangle |^2
      \over
      E_{u , n} - E_u };
  \end{equation}
  $\Phi_{u , n }$, $E_{u , n }$ are the $n$-th eigenstate and energy eigenvalue
  of the effective Hamiltonian $H_u$.

  The second-order energy difference $T_u[w]$ is a functional of both the potentials $u$ and $w$, but for now we take $u$ to be fixed and focus on its dependence on $w$. In the following, we seek to minimize $T_u[u+v']$ over $v'$: this is equivalent to minimizing $T_u[w]$ over $w$, because $w=u+v'$ and $u$ is fixed.
  In~\cite{gidopoulos1} the same symbol $v$ was used for the
  potential appearing as the argument of the functional $T_\Psi $ in~\eqref{eq1}
  and for the argument of $T_u$ in~\eqref{eq8}.
  Here, we use different symbols $v$ and $w$ to avoid confusion.

  The functional derivative of $T_u [w]$ with respect to $w$, at fixed $u$, is given by\footnote{Note that in~\cite{gidopoulos1}, the functional derivative in Eq.~\eqref{eq9} has the wrong sign. It is correct in this paper.}
  \begin{eqnarray}
    \lefteqn{ {\delta T_u[w] \over \delta w ({\bf r}) } = } \nonumber \\
    && \sum_{i, \, a } 
            { \langle \phi_{u,i} | {\cal J}_u  - {\cal K}_u - { w}  | \phi_{u,a} \rangle
              \over {\epsilon_{u , i} - \epsilon_{u , a}} } \, \phi_{u , a}^*  ({\bf r})  \phi_{u , i} ({\bf r})
            + {\rm c.c.} \ \label{eq9}
  \end{eqnarray}
  ${\cal J}_u({\bf r})$ is the direct Coulomb (or Hartree) local potential operator
  and ${\cal K}_u$ is the Coulomb exchange non-local operator. $\phi_{u , i}$ and $\phi_{u , a}$ are respectively
  occupied and unoccupied orbitals in the Slater determinant $\Phi_u$, with $\epsilon_{u , i}$ and $\epsilon_{u , a}$ their corresponding eigenvalues.
  The functional derivative in Eq.~(\ref{eq9}) represents a charge density with zero net charge,
  \begin{equation} \label{eq10}
    \int d{\bf r} \,  { \delta T_u [w] \over \delta w ({\bf r}) } = 0 .
  \end{equation}

  Optimization over $w$ in Eq.~(\ref{eq8}), by setting the functional derivative (\ref{eq9}) equal to zero (at fixed $u$),
  yields the first order KS potential. We denote by $w_0 [u]$ the minimizing potential of $T_u[w]$
  for fixed $u$,
  \begin{equation} \label{eq11}
    \min_w T_u [w ] = T_u \big[w_0 [u] \big] .
  \end{equation}
  From (\ref{eq7}), the first-order term $v'[u]$ in the KS expansion can be obtained from
  \begin{equation} \label{eq12}
    w_0 [u] ({\bf r})  =  u({\bf r}) +  v' [u] ({\bf r})  .
  \end{equation}
  The desired expansion of the KS potential to first order is (\ref{eq2},\ref{eq6})
  \begin{equation} \label{eq13}
    v_s [u] ({\bf r}) = v_{en}({\bf r}) + u({\bf r}) + \alpha \, v' [u] ({\bf r}) + \mathcal{O}(\alpha^2).
  \end{equation}

  The exact KS potential does not depend on $u$, but when the expansion is truncated at a finite order,
  the KS potential up to that order will depend on $u$.
  Hence, we write $v_s[u]$ to denote the KS potential up to first-order, and $v_s$ to denote the exact KS potential.
  We also denote by $\Phi_s[u]$ the g.s. of $v_s[u]$, i.e.\ the KS determinant
  of the first-order KS potential $v_s[u]$.

  In the Taylor expansion of the KS potential (\ref{eq13}), the zeroth-order term, $ v_{en}({\bf r}) + u({\bf r})$, is the same as the potential in $H_u$.
  The first-order term in the expansion of $v_s $ is $v' [u] $.
  We may construct as many expansions for the KS potential as there are choices
  for $u$, and more besides using an altogether different expansion for $\Psi$, such as
  M{\o}ller-Plesset.

  It is interesting to note that, by setting $w=u$ in the functional derivative~\eqref{eq9}, we retrieve the equation for the exchange-only
  OEP (xOEP), also known in the literature as (exchange-only) exact exchange potential (EXX).
  This particular choice of $u$ will be discussed in more detail in section 5; for now, we see how it also arises from an alternative
  perspective.

  The density $\rho_{\Psi_u(\alpha)} ({\bf r}) $ of the weakly interacting state $\Psi_u ( \alpha )$ is given by
  \begin{equation} \label{eq14}
    \rho_{\Psi_u(\alpha)} ({\bf r})  = \rho_u ({\bf r}) \, + \, \alpha \, \left. {\delta T_u[w] \over \delta w ({\bf r}) } \right|_{ w = u}
    \!\!\! + \mathcal{O}( \alpha^2 ),
  \end{equation}
  where $\rho_u ({\bf r})$ is the density of the zeroth-order state $\Phi_u$.
  The density $\rho_{\Psi_u(\alpha)} ({\bf r}) $ of the weakly interacting system differs from the zeroth-order
  density $\rho_u({\bf r})$ by a charge density equal (up to first order) to the functional derivative (\ref{eq9}), where the latter is
  evaluated at $w = u$.
  Therefore, the search for the zeroth order potential $u$ for which the g.s. density does not change to first order yields the exchange-only OEP (xOEP),
  as observed by Bartlett and coworkers~\cite{bartlett_collapse}.

  Furthermore, the density $\rho_{\Psi_u(\alpha)}$ is related to the density $\rho_{u+\alpha v'}\r$ as follows:
  \begin{equation} \label{eq15}
    \rho_{\Psi_u(\alpha)} ({\bf r}) =  \rho_{u+\alpha v'} ({\bf r}) + \alpha \, 
    \left. {\delta T_u[w] \over \delta w ({\bf r}) } \right|_{ w = u+v'}
    \!\!\! + \mathcal{O}( \alpha^2 ).
  \end{equation}
  Hence, the density $\rho_{\Psi_u(\alpha)}$ of the weakly interacting state differs from the density $\rho_{u+\alpha v'}\r$ of the non-interacting state by a charge density equal (up to first order) to the functional derivative~\eqref{eq9}, where the latter is evaluated at $w=u+v'$. Therefore, these densities are equal if the potential $w$ is equal to the minimizing potential $w_0[u]$~\eqref{eq12}; this minimizing potential defines the KS potential $v_s[u]$~\eqref{eq13}. In other words, for any $u$, the density of the KS state is equal to the density of the weakly-interacting state (to first order),
  \begin{equation} \label{eq16}
    \rho_{s} [u] \r = \rho_{\Psi_u(\alpha)}\r + \mathcal{O}(\alpha^2),
  \end{equation}
  where $ \rho_{s} [u] \r = \rho_{u+\alpha v'[u]}\r$.

  Although there can be several $u$ that yield a converging expansion for $\Psi$ and for
  the KS potential $v_s$, we want to find those $u$ whose expansions converge faster than others. 
  We investigate this in section
  \ref{sec:3}.
  \subsection{Relation with the Sham-Schl\"uter method}
  
  Before proceeding to section \ref{sec:3}, we make contact with MBPT and the formalism of 
  Green's functions. In MBPT, the requirement by Kohn and Sham that the density of the auxiliary noninteracting (KS) 
  system be equal to the density of the interacting system leads to the Sham-Schl\"uter equation \cite{sham_schluter,godby_sham_schluter,gross_book},
  \begin{multline}
 \label{sham_schluter}
 \int d {\bf r} ' v_{xc} ({\bf r} ' ) \int d \omega \, G_s ({\bf r} , {\bf r} ' ; \omega ) \, G  ({\bf r} ' , {\bf r} ; \omega ) \\
 = \iint d {\bf x} d {\bf y} \int d \omega \, G_s ({\bf r} , {\bf x} ; \omega ) \, \Sigma _{xc} ({\bf x} , {\bf y} ; \omega )
  \, G  ({\bf y} , {\bf r} ; \omega ) \, , 
   \end{multline}
   in which $G  ({\bf r}  , {\bf r} ' ; \omega )$ and $G_s  ({\bf r} , {\bf r} ' ; \omega )$ are respectively the one-particle Green's functions for the 
   interacting and the noninteracting (KS) systems.  
   Eq.~\eqref{sham_schluter} determines the approximate exchange and correlation (xc) potential $v_{xc}$ in terms of 
   an approximate xc self-energy $\Sigma_{xc} ({\bf r} , {\bf r} ' , \omega)$.  
  
  Following Engel and Dreizler \cite{engel_dreizler}, who derive the OEP equation for the xc potential from the Sham-Schl\"uter 
  equation, we point out the relation between Eq.~\eqref{sham_schluter} and Eqs. (\ref{eq15}) and (\ref{eq16}). 
  Using \eqref{eq15} and requiring that the densities of the noninteracting and interacting systems be equal up to first order, i.e. 
  requiring the validity of \eqref{eq16}, yields the OEP equation,
  \begin{equation} \label{eq18}
  \left. {\delta T_u[w] \over \delta w ({\bf r}) } \right|_{ w_0 [u]} = 0 ,
  \end{equation}
  which determines the first-order KS potential $v ' [ u ] $ \eqref{eq12}. This equation for $v'[u]$ is equivalent to the Sham-Schl\"uter equation 
\eqref{sham_schluter} with $v'[u]$ in place of $v_{xc}$ and the modified self-energy, $\Sigma - u$, in place of $\Sigma_{xc}$.   

 Of course, in our theory, we do not impose the validity of \eqref{eq16}, since the equality of the two densities 
 comes out naturally from the optimisation of the second-order energy difference $T_u [ w ]$ \eqref{eq8}. 


  \section{Reference determinants with minimum correlation energy} \label{sec:3}

  Historically, the xOEP is found by a minimization of the total
  energy
  $\langle \Phi_v | H | \Phi_v  \rangle$, where the Slater determinant $\Phi_v$ depends on
  the effective potential $v({\bf r})$ (see Eq.~\ref{eq2}).
  Since the exact energy $\langle \Psi | H | \Psi \rangle$ does not depend on $v$, the minimization of the energy is
  equivalent to the minimization over $v$ of the magnitude of the correlation energy from the reference Slater determinant $\Phi_v$,
  \begin{equation} \label{eq19}
    E_H ^c [ v ] \doteq  \langle \Psi | H | \Psi \rangle - \langle \Phi_v | H | \Phi_v  \rangle < 0;
  \end{equation}
  we have explicitly shown the dependence of the correlation energy on the interacting Hamiltonian $H$ of the system and on $v$.
  Hence, another interpretation of the xOEP follows:
  \begin{corollary}
    xOEP is that effective potential $v ({\bf r})$ with weakest correlation energy from its ground state $\Phi_v$.
  \end{corollary}

  The implication is that if we want to treat the interacting Hamiltonian perturbatively to all orders, then
  the effective Hamiltonian with the xOEP potential is the best zeroth-order
  Hamiltonian, as the remaining correlation energy to be treated perturbatively
  is smallest.


  Often, we are interested in the lowest orders of perturbative expansions
  either because we want to study the limit of weak interactions or
  because we can only access the lowest orders numerically.
  Hence, we consider the partially interacting system described by the
  perturbative Hamiltonian $H_u (\alpha)$ in (\ref{4}) where the zeroth-order
  potential $u ({\bf r})$ is meant to be determined later on in an optimal way.
  We make the following statement for the weakly interacting system described by
  the Hamiltonian $H_u (\alpha)$, in the limit $\alpha \rightarrow 0$ and for any
  $u$:
  \begin{lemma}
    The KS potential $v_s[u] ({\bf r})$ is that effective potential with weakest correlation
    energy from its ground state $\Phi_s[u]$.
  \end{lemma}
  In this statement, the KS potential $v_s[u]$ is given to first order and the
  lowest (dominant) order in the correlation energy is second.

  \begin{proof}

    The correlation energy for the partially interacting system using as reference
    the g.s. $\Phi_v$ (see Eq.~\ref{eq3}) of an effective local potential $v({\bf r})$
    in the class of Hamiltonians (\ref{eq2}) is:
    \begin{equation} \label{eq20}
      E_{H_u (\alpha)} ^c [ v ] \doteq
      E_u (\alpha) - \langle \Phi_v | H_{u} (\alpha) | \Phi_v  \rangle  < 0 .
    \end{equation}
    %
    For fixed $u$, the potential that minimizes the magnitude of the
    correlation energy $E_{H_u (\alpha)} ^c [ v ]$ over $v$ is the same as the
    potential that minimizes the expectation value \\
    $\mel*{\Phi_v}{H_u(\alpha)}{\Phi_v}$ over $v$, since $E_u (\alpha)$ does not
    depend on $v$. This optimal effective potential is different in general
    from the xOEP, due to the dependence of the former on $u$ and on $\alpha$.

    Let us expand the correlation energy~\eqref{eq20} in powers of
    $\alpha$ and obtain the dominant term.
    Obviously, when $\alpha = 0$, the potential $v$ that minimizes the energy
    $\langle \Phi_v | H_{u} ( 0 ) | \Phi_v  \rangle$
    (or minimizes the magnitude of $E_{H_u ( 0 )} ^c [ v ]$) is $v=u$.
    Hence, for small $\alpha$, we substitute Eq.~\ref{eq6} in (\ref{eq20}) and we expand the correlation energy
    \begin{equation} \label{eq21}
      E_{H_u (\alpha)} ^c [ u+\alpha v' ] \doteq E_u( \alpha) - \langle \Phi_{u + \alpha v'} | H_{u} (\alpha) | \Phi_{u + \alpha v'} \rangle
    \end{equation}
    to second order in $\alpha$ to obtain
    \begin{equation} \label{eq22}
      E_{H_u (\alpha)} ^c [ u + \alpha v' ] = - \alpha^2 T_u [u + v'] +
      {\cal O}(\alpha^3)  ,
    \end{equation}
    where $T_u [w]$ is given by (\ref{eq8}).

    Up to second order in $\alpha$, the correlation energy (\ref{eq22}) is thus equal to minus the energy difference (\ref{eq7}):
    \begin{equation} \label{eq23}
      E_{H_u (\alpha)} ^c [ u + \alpha v' ] = - T_{\Psi_u (\alpha)} [ u + \alpha v' ] .
    \end{equation}
    The KS potential $v_s[u]$ in~\eqref{eq13} is that potential which minimizes the energy difference, and hence the statement follows.
  \end{proof}

  It follows that when we minimize $T_u [w]$ over $w$ to obtain the first order KS potential $v_s[u]$, the resulting potential
  not only has the same density as  $\Psi_u (\alpha)$ to first order, but it also has the following unique properties among other effective local potentials:
  \begin{itemize}
  \item it best adopts $\Psi_u (\alpha)$ (to first order) as its own approximate ground state and
  \item its KS ground state $\Phi_s [u]$ has the lowest magnitude of correlation energy to second order.
  \end{itemize}

  Let us denote by $E^c_u [w]$ the negative of the energy difference $ T_u [w]$,
  \begin{equation} \label{17b}
    E^c_u [ w ]  =  - T_u [w].
  \end{equation}
  $E^c_u [w]$ is a second order correlation energy expression.
  It is useful to use this notation to represent the total energy of the
  weakly interacting systems described by $H_u ( \alpha)$ using three different
  references: the zeroth-order state $\Phi_u$, the perturbative state $\Phi_{u + \alpha v'}$, and the KS determinant $\Phi_s[u]$. Keeping up to second order, we have
  in the limit $\alpha \rightarrow 0$:
  \begin{align}
    \lefteqn{E_u (\alpha) =  \langle \Phi_u | H_{u} (\alpha) | \Phi_u  \rangle+  \alpha^2 E^c_u [ u ]  + {\cal O}(\alpha^3) } \label{kosi} \\
    &= \langle \Phi_{u + \alpha v'} | H_{u} (\alpha) | \Phi_{u + \alpha v'}  \rangle + \alpha^2 E^c_u [ u +  v' ]  + {\cal O}(\alpha^3) \label{kosiena} \\
    &= \langle \Phi_s[u] | H_{u} (\alpha) | \Phi_s[u]  \rangle + \alpha^2 E^c_u \big[ u + v'[u] \big]  + {\cal O}(\alpha^3) ,
    \label{kosidyo}
  \end{align}
  where $\Phi_s[u]$ is the ground state of the first order KS potential $v_s[u]$ in~\eqref{eq13}.
  In general, for a given $u$, the optimal potential $v'[u] $ (\ref{eq12}) does not vanish
  and therefore the first-order KS potential $v_s[u]$ is different to the zeroth-order
  potential $u$.

  In the following, we shall determine $u$ optimally by selecting the one that makes
  $E^c_u \big[ u + v'[u] \big]$ small.

  \section{Comparison of DFT PT and the present WFT}


  \subsection{Traditional DFT PT method}

  In traditional DFT PT the KS potential is obtained from a perturbative expansion of the total energy functional, thus they are of the same order. The first order term in the expansion of the total energy is the exact exchange energy functional, which yields through functional differentiation
  the exchange potential, the first order term in the expansion of KS potential.
  Similarly, the correlation energy functional (truncated at second order) yields the correlation potential, which
  is the second order term in the expansion of the KS potential.
  The familiar scheme is summarized below:

  
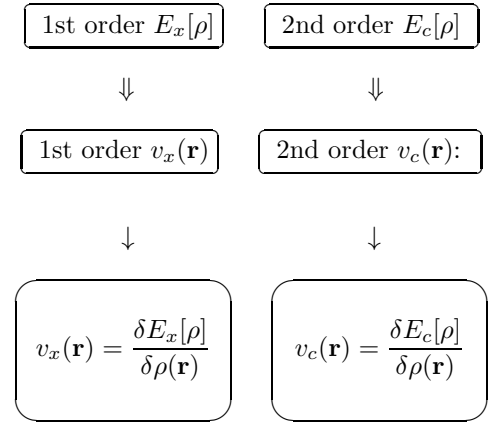
\begin{center}
    \vbox{
      \vspace{0.5cm}

      \ovalbox{ 1st order $E_x [\rho]$} \hspace{0.3cm} \ovalbox{ 2nd order $E_{c} [\rho]$ }
      $$
      \Downarrow \hspace{3.1cm} \Downarrow
      $$

      \ovalbox{ 1st order $v_x ({\bf r})$}  \hspace{0.3cm} \ovalbox{ 2nd order $v_{c} ({\bf r})$: }

      $$
      \downarrow \hspace{3.1cm} \downarrow
      $$
      \ovalbox{
        \begin{minipage}{2.4cm}
          $$
          v_{x} ({\bf r}) = {\delta E_{x} [\rho]
            \over \delta  \rho ({\bf r})}
          $$
          $ $
        \end{minipage}
      }
      \hspace{0.3cm}
      \ovalbox{
        \begin{minipage}{2.4cm}
          $$
          v_{c} ({\bf r}) = {\delta E_{c} [\rho]
            \over \delta  \rho ({\bf r})}
          $$
          $ $
        \end{minipage}
      }
    }
  \end{center}
  \vspace{0.5cm}

  Because the exact exchange energy cannot be written explicitly in terms of the density, its functional derivative (the exact exchange potential)
  cannot be obtained directly from the density but only after solving an integral equation (Fredholm equation of the first kind), 
  known as the equation for the optimized effective potential method.
   Although we are solving an OEP equation, the exchange potential is still 
    the functional derivative of the exchange energy functional w.r.t. the density.

  \subsection{Present WFT method} \label{sec:4.2}

  In the present WFT method, which happens to have the KS potential as its solution, the
  xc potential is not the functional derivative of the xc energy w.r.t.\ the density
  (since the
  various quantities are not density functionals) and cannot be obtained directly.
  Instead, minimization of the magnitude of the \emph{second-order} correlation energy
  functional $E_u^c [ v ]$,
  Eq.~\eqref{17b}, yields the minimizing potential $w_0 [u]$,
  which emulates the Hartree exchange and correlation potential (Hxc) 
  for the KS system with density $\rho_s[u]$ \eqref{eq16}.
  The sum $v_{\text{en}} + w_0 [u] $ gives the KS potential up to \emph{first order},
  Eq.~\eqref{eq13}, for $\alpha=1$.

  The xc-potential term in $v_s[u]$ is obtained by subtracting the Hartree potential from
  the optimal potential $w_0[u]$.
  The scheme is summarized below:
  
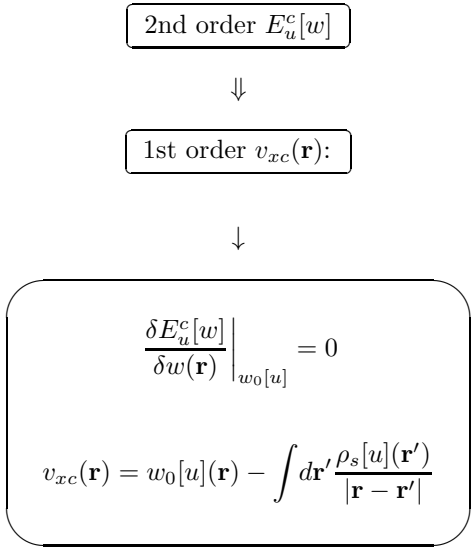
\begin{center}
    \vbox{
      \vspace{0.5cm}

      \ovalbox{$ $ 2nd order $E_u^c [w ]$ $ $}
      $$
      \Downarrow
      $$

      \ovalbox{$ $ 1st order $v_{xc} ({\bf r})$: $ $}

      $$
      \downarrow
      $$
      \ovalbox{
        \begin{minipage}{5.7cm}
         
          $$
          \left. {\delta E_u^c [ w ]
            \over \delta w ({\bf r})} \right|_{w_0 [u]} = 0
          $$
          $$
          v_{xc} ({\bf r}) = w_0 [u] ({\bf r}) - \int \! d {\bf r}' { \rho_s[u]({\bf r}') \over |{\bf r} - {\bf r}' |}
          $$
          $ $
        \end{minipage}
      }
    }
  \end{center}
  \vspace{0.5cm}

  We emphasize again the conceptual shift between the two theories: in DFT PT, the KS potential is obtained by minimizing the total energy of the system,
  while in the present WFT method the KS potential is obtained by minimizing the energy difference $T_\Psi [v]$.
  To dominant order, the latter optimization amounts to minimizing the magnitude of the correlation energy from the KS determinant.

  \section{Optimal choices for $u$}

  In the following we shall explore some choices for approximations to the interacting state $\Psi$.
  Based on the expansion $\Psi_u(\alpha)$ discussed so far, this amounts to making a suitable
  choice for the potential $u$.
  However, we are free to pick any $\tilde\Psi$ which might be expected to yield an accurate
  approximation to the exact KS potential;
  in addition, we shall also consider a M{\o}ller-Plesset expansion for $\Psi$. In any case,
  since we shall only consider perturbative expansions for $\Psi$, we wish to find expansions for the KS potential $v_s$
  which are expected to give accurate results when the expansion is truncated at the lowest (meaningful) order: first order
  for $\Psi$ and $v_s$, and second order for the correlation energy.

  In traditional DFT PT, the first order KS potential is restricted to Hartree and exact exchange,
  in fact in DFT PT the Hartree and exchange potential is {\em defined\/} as the first order term in the expansion of
  the KS potential.
  We shall discuss two choices for $H_u$ for which the first order KS potential indeed corresponds to Hartree and
  exchange only.
  Finally, we shall introduce a third choice for $H_u$, which is expected to yield a first order KS potential with accurate
  Hartree and exchange and correlation character.

  \subsection{Exchange optimized effective potential} \label{sec:5.1}

  We anticipate that a good choice for $u$ is such that the magnitude of the second-order correlation energy 
  $\big| E_u^c[u] \big| = T_u [ u ]$ (\ref{17b},~\ref{kosi}) is small,
  but we shall not discuss here how to find the global minimum of $T_u [ u ]$. 
  An energetically better choice will be investigated in section~\ref{CC}.
  However, we present below an alternative argument which allows us to pick a $u$ for which $T_u [ u ]$ is small. 
  That choice of $u$ yields xOEP.

  For all zero-order potentials $u$, it holds that:
  \begin{equation} \label{29}
    \min_{w} T_u [ w ] \le T_u [ u ]
  \end{equation}
  The inequality holds because the search for the minimum over the potential $w$ includes
  the value $ w = u$.
  Inequality (\ref{29}) states that for any potential $u$, the magnitude of its
  correlation energy
  $\big| E_u^c[u] \big| = T_u [ u ] $ is always larger or at most equal to the minimum of
  $T_u [ w ]$. It follows that
  the potential $u_{Hx}$, for which equality holds in~\eqref{29},
  \begin{equation} \label{30}
    T_{u_{Hx}} \big[ u_{Hx} + v'[u_{Hx}] \big] = T_{u_{Hx}} [ u_{Hx} ] ,
  \end{equation}
  will have correlation energy with small magnitude (but not the smallest possible).
  Equality in~\eqref{30} holds when the first-order term in the expansion of the KS potential vanishes,
  \begin{equation} \label{28b}
    v'[u_{Hx}] = 0 \, .
  \end{equation}
  The potential $u_{Hx}$ is then determined by setting $w = u_{Hx}$
  in Eq.~\eqref{eq9} and finding the potential $u_{Hx}$ which makes this functional
  derivative vanish,
  \begin{multline} \label{eq:17}
    \sum_{i,a}\frac{\mel{\phi_{{u_{Hx}},i}}{J_{{u_{Hx}}}-K_{{u_{Hx}}}-{u_{Hx}}}{\phi_{{u_{Hx}},a}}}{\epsilon_{{u_{Hx}},i}-\epsilon_{{u_{Hx}},a}}\times \\
    \phi^*_{{u_{Hx}},a}\rvec{r}\phi_{{u_{Hx}},i}\rvec{r}+\cc = 0
  \end{multline}
  This is the well-known equation for the xOEP.
  Hence the KS potential is
  \begin{eqnarray}\label{KS_Hx}
    v_s[u_\text{Hx}]\r=v_\text{en}\r + u_\text{Hx}\r +\mathcal{O}(\alpha^2).
  \end{eqnarray}

  We note two differences between our method and DFT PT, which also yields the xOEP:
  \begin{enumerate}[(a)]
  \item In DFT PT the xOEP is the functional derivative of the exchange energy functional, which appears as the first-order term in the DFT PT expansion of the xc energy functional.
  \item In DFT PT, the total energy that gives rise to xOEP is truncated to first order and includes exchange energy and no correlation energy. There is no way to pair xOEP with a correlation energy functional without self-consistently
    altering the exchange potential away from its exchange only character.

    In the current WFT, the first order KS potential is always paired, naturally, with a second
    order correlation energy, even when the first-order potential is xOEP.
    Specifically, the correlation energy corresponding to xOEP is given by $E^c_{u_{Hx}}[u_{Hx}]=-T_{u_{Hx}}[u_{Hx}]$.
  \end{enumerate}

Finally, we remark that the xOEP can be obtained \cite{sham_schluter} from the Sham-Schl\"uter equation 
\eqref{sham_schluter}, when we keep only the exchange term in the self-energy and approximate the interacting  Green's function $G$ 
with $G_s$ (linear Sham-Schl\"uter equation).

  \subsection{Local Fock exchange potential}

  So far, we have approximated the interacting state $\Psi$ with the partially interacting state $\Psi_u(\alpha)$, and considered which local potentials $u\r$ will give accurate approximations to the KS potential. In the prior section, we saw how one particular choice of $u$ yields the well-known xOEP. However, we now consider an altogether different approximation to $\Psi$, the M{\o}ller-Plesset (MP) expansion $\Psi_\text{MP}$.

  We initially consider only the zeroth-order term in the MP expansion, the Hartree-Fock (HF) determinant $\Phi_\text{HF}$. Following the approach in~\cite{LFX_Hollins}, we search for the effective Hamiltonian $H_v$, with local potential $v\r$
  (Eq.~\ref{eq2}),
  which optimally adopts $\Phi_\text{HF}$ as its ground-state. We therefore minimize the energy difference $T_\text{HF}[v]$, given by
  \begin{equation}
    T_\text{HF}[v]=\mel{\Phi_{\text{HF}}}{H_v}{\Phi_{\text{HF}}} -E_v,\label{28}
  \end{equation}
  over $v\r$ to determine the optimal $H_v$. The functional derivative of  $T_\text{HF}[v]$ is equal to
  \begin{equation} \label{34}
    \fdv{T_\text{HF} [v]}{v\r}= \rho_{\text{HF}}\r -\rho_v \r.
  \end{equation}
  where $\rho_{\text{HF}}\r$ is the HF density, i.e.\ the density of $\Phi_{\text{HF}}$.

  The ground-state whose potential minimizes this energy difference thus has the same density as the HF determinant. Denoting this optimal potential as $v_\text{MP0}$, the local Fock-exchange (LFX) potential is defined as\footnote{This is defined differently in~\cite{LFX_Hollins} because in the current paper we do not include the electron-nuclear potential $v_\text{en}$ in the local potential $v$.}
  \begin{equation}
    v_\text{LFX}\r = v_\text{MP0}\r -\int\dd{\vec{r}'}\frac{\rho_\text{HF}(\vec{r}')}{|\vec{r}-\vec{r}'|};
  \end{equation}
  and the MP expansion of the KS potential is
  \begin{equation}
    v_s^\text{MP} \r = v_\text{en}\r +\int\dd{\vec{r}'}\frac{\rho_\text{HF}(\vec{r}')}{|\vec{r}-\vec{r}'|}+v_\text{LFX}\r +\mathcal{O}(\alpha^2).
  \end{equation}
  The local potential with the HF density has been considered previously in the literature as an accurate approximation to xOEP and EXX
  \cite{levy_pnas,payne_1979,staroverov2013}.
  Much like for $\Psi_u(\alpha)$ and subsequent expansions of the KS potential $v_s[u]$, we can consider higher order terms in the MP expansion of
  $\Psi_{\rm MP}$, which give rise to MP expansions of the KS potential. However, from Brillouin's theorem~\cite{szabo2012modern}, singly excited Slater determinants do not couple directly with their zeroth-order HF state, which means the density of $\Psi_\text{MP}$ does not change to first order. Therefore, the potential which optimizes the energy difference,
  \begin{equation}
    T_{\Psi_\text{MP1}}[v]=\mel*{\Psi_{\text{MP}1}}{H_v}{\Psi_{\text{MP}1}} - E_v,
  \end{equation}
  where $\Psi_{\text{MP}1}$ is the first-order MP state, is the same potential as that
  which minimizes the energy difference in Eq.~\eqref{28}. Including first-order corrections to the MP expansion thus leaves the
  density and hence the expansion of the KS potential unchanged up to first-order.
  This is entirely analogous to our derivation of the xOEP:
  the xOEP is that zero-order effective potential ($u_{Hx}$), such that when we switch on the Coulomb interaction, the g.s. density of the weakly
  interacting state does not change to first-order~\eqref{eq14}, and whose corresponding power series expansion for the KS potential
  also has vanishing first-order correction~\eqref{28b}. 

  Both potentials (LFX and xOEP) have exchange character.
  In DFT PT, the xOEP is the exact exchange potential (EXX) as it is the functional derivative of the exchange energy functional with respect
  to the density.
  The LFX potential cannot be expressed exactly but only approximately~\cite{LFX_Hollins} as the functional derivative of the exchange
  energy functional.
  
  Similarly to the xOEP, the LFX potential as well can be obtained from the Sham-Schl\"uter equation \eqref{sham_schluter} 
when we keep the exchange part of the self-energy and omit correlation \cite{gross_book}. 
However, unlike xOEP, the linear-response approximation (the replacement of $G$ by $G_s$) is not employed to determine the LFX potential, 
and hence, from the point of view of the Sham-Schl\"uter method, the xOEP is an approximation of the LFX potential. 
On the other hand, from the DFT point of view, the LFX potential is instead an approximation of the exact exchange 
potential since only the latter is the functional derivative of the exact exchange energy w.r.t. the density \cite{staroverov2013,LFX_Hollins}.

  As discussed in~\cite{LFX_Hollins}, the LFX and xOEP potentials are mathematically distinct, but share many physical properties, and
  would thus be expected to yield similar results when exchange dominates over correlation. Indeed, this was demonstrated to be the case,
  and it was theorized that the difference in results between the two methods is likely to indicate the correlation strength for a given system.
  Although the two methods are very similar, one advantage of the LFX method is that the functional derivative~\eqref{34} is easier to
  compute, as there is no need to calculate the KS orbital shifts~\cite{OEP_Hylleraas,OEP_Kummel}.

  \subsection{First order exchange and correlation potential}\label{CC}

  We previously saw that making the magnitude of the correlation energy $E_u^c[u] $~\eqref{kosi} small gave rise to the well-known
  Hartree and exact exchange potential in the first order KS potential~\eqref{KS_Hx}. Whilst it is interesting to reproduce this result with our
  method, we want to develop a new expression that will give accurate results for systems where correlation is important.

  As mentioned, finding the absolute minimizing potential of $\big| E_u^c[u] \big|$~\eqref{kosi} is mathematically complex.
  The argument which gave rise to the Hartree and exact exchange potential does not fully optimize
  $\big| E_u^c[u] \big|$,
  and thus the expansion of $v_s[u]$ is not expected to converge as fast as desired.
  Let us instead try to minimize the magnitude of the correlation energy $E_u^c[u+v']$.
  In principle, this involves a coupled minimization over $u$ and $v'$ which is even more complicated than minimizing $\big| E_u^c[u] \big| $
  over $u$.
  However, in practice the two minimizations of $E_u^c[u+v']$ can be approximately decoupled, simplifying significantly the minimization scheme.
  To proceed, we split $T_u[w]$ into two terms,
  \begin{equation} \label{17}
    T_u [ w ] = S_u[w] + D [u],
  \end{equation}
  with
  \begin{eqnarray}
    S_u [ w ] & = & \sum_{n \  {\rm single}} { | \langle \Phi_{u , n } | V_{ee} - \sum_i w ({\bf r}_i ) | \Phi_u \rangle |^2
      \over
      E_{u , n} - E_u } \label{18}
  \end{eqnarray}
  and
  \begin{eqnarray}
    D [u] & = & \sum_{n \ {\rm double}} { | \langle \Phi_{u , n } | V_{ee} | \Phi_u \rangle |^2
      \over
      E_{u , n} - E_u }. \label{19}
  \end{eqnarray}
  The first term $S_u[w]$ is a sum is over singly excited determinants from $\Phi_u$,
  while the second term $D[u]$ is a sum over doubly excited determinants.

  The potential $w$ appears only in $S_u[w] $ but not in $D[u]$. Hence the
  minimizing potential $w_0 [u]$ of $T_u [w]$ also minimizes $S_u[w] $ but leaves $D[u]$
  unaffected.

  In practice~\cite{Gido_Lath}, we have found that for any reasonable $u$, 
  the minimization of $T_u [w]$ over $w$ reduces $S_u [w]$
  to very small values, compared with $D[u]$:
  \begin{equation} \label{20}
    0 <   S_u \big[ w_0 [u] \big] \ll D[u] .
  \end{equation}
  Therefore, the dominant term is $D[u]$, and the minimum of the energy difference $T_u[w]$ over $w$ is given by $D[u]$ to a good
  approximation,
  \begin{equation} \label{21}
    T_u \big[ w_0 [u] \big] \simeq D[u].
  \end{equation}
  We conclude that in order to pick the best $u$, so as to minimize the minimum $T_u \big[ w_0 [u] \big]$,
  it is sufficient to choose $u$ to minimize the double-excitations term $D[u]$.
  This optimal $u_0$, with minimum $D [u_0]$ will correspond to the best zeroth-order effective
  Hamiltonian $H_{u_0}$ for a perturbative expansion of the interacting Hamiltonian (\ref{4}). This dominance of $D[u]$ over $S_u[w_0[u]]$ also
  reinforces that $u=u_{Hx}$ is not energetically the best choice of $u$, since $D[u]$ is not optimised in
  any way
  by this choice and can be quite large.

  To minimize the double-excitations term, we first need to derive the functional derivative of $D[u]$,
  \begin{equation}\label{eq:fdv}
    \int \dr{r} \, \delta u\rvec{r} \, \fdv{D[u]}{u\rvec{r}} = \lim_{\lambda\to 0}\frac{D[u+\lambda \delta u]-D[u]}{\lambda},
  \end{equation}
  so we need to determine how $D[u]$ changes due to a perturbation $u\to u+\lambda \delta u$. Given that the ground and excited state wavefunctions, $\Phi_u$ and $\Phi_{u,n}$, as well as their respective energy levels, $E_u$ and $E_{u,n}$, are affected by the perturbation, $D[u+\lambda  \delta u]$ to first order is
  \begin{equation}\label{eq:DU1}
    D[u+\lambda \delta u]=
    \sum \frac{\big| \vmel{\Phi_{n}+\lambda\Phi_{\delta u,n}^{(1)}}{\Phi_0+\lambda \Phi_{\delta u,0}^{(1)}}\big|^2}{E_{n}+\lambda E_{\delta u,n}^{(1)}-E_0-\lambda E_{\delta u,0}^{(1)}},
  \end{equation}
  where the dependence on $u$ is now assumed and $\Phi_0$ labels the ground state. We use the notation $\vmel{\Phi_1}{\Phi_2}=\mel{\Phi_1}{V_{ee}}{\Phi_2}$. To write $D[u+\lambda \delta u]$ explicitly to first order in $\lambda$, we multiply it by the denominator in Eq.~\eqref{eq:DU1} and then write both sides of the subsequent expression as a power series in $\lambda$. We then expand the squared term which yields the following expression for the r.h.s.\ of Eq.~\eqref{eq:fdv},
  \begin{align}
    &\lim_{\lambda\to 0}\frac{D[u+\lambda \delta u]-D[u]}{\lambda}=
    \sum\frac{\vmel{\Phi_{n}}{\Phi_u}}{E_{n}-E_0}
    \bigg\{\vmel{\Phi_0}{\Phi_{\delta u,n}^{(1)}}\nonumber\\
    &+\vmel{\Phi_{\delta u,0}^{(1)}}{\Phi_{n}}
    -\frac{1}{2}\frac{E^{(1)}_{\delta u,n}-E^{(1)}_{\delta u,0}}{E_{n}-E_0}\vmel{\Phi_0}{\Phi_{n}}\bigg\}+\cc\label{eq:bigsum}
  \end{align}

  We must now determine the perturbed states and energies. We begin with the perturbed state $\ket*{\Phi^{(1)}_{\delta u,n}}$; from
  Rayleigh-Schr\"odinger perturbation theory, this is given by
  \begin{equation}\label{eq:PT1}
    \ket*{\Phi_{\delta u,n}^{(1)}}=\sum_{m\neq n}\frac{\mel*{\Phi_m}{\delta U}{\Phi_n}}{E_n-E_m}\ket*{\Phi_m} ,
  \end{equation}
  where $\delta U \doteq  \sum_i \delta u ({\bf r}_i)$.

  Since $\ket*{\Phi_n}$ is a doubly excited state, we can write it in the form $\ket*{\Phi_{ij}^{ab}}$, where $i,j$ denote occupied orbitals in the ground state and $a,b$ denote unoccupied orbitals. The matrix element, $\ket*{\Phi_{\delta u,n}^{(1)}}$, is evaluated using Slater-Cordon rules and is given by
  \begin{equation}
    {\ket*{\Phi_{\delta u,n}^{(1)}}}=\sum_{c}\sum_{k}\frac{\mel*{c}{\delta u}{k}}{\epsilon_k-\epsilon_c}\ket*{\Phi_{ijk}^{abc}},
  \end{equation}
  where $k\in\ket*{\Phi_{ij}^{ab}}$, and $c\not\in\ket*{\Phi_{ij}^{ab}}$. The possible combinations for the pair $(k,c)$ are therefore
  \begin{equation}
    (a,i);(a,j);(b,i);(b,j);(\mu,i);(\mu,j);(a,\nu);(b,\nu),
  \end{equation}
  where $\mu\neq(i,j),\ket{\mu}\in\ket{\Phi}$ and $\nu\neq(a,b),\ket{\nu}\not\in\ket{\Phi}$. Any other permissible combination of $(k,c)$ represents a triple excitation which will vanish in the final expression. We now determine the state $\ket*{\Phi_{ijk}^{abc}}$ based on these possible combinations. We write
  \begin{equation}
    \ket*{\Phi_{ijk}^{abc}}=\create_c\destroy_k\create_b\destroy_j\create_a\destroy_i\ket{\Phi},
  \end{equation}
  where $\create$ and $\destroy$ are fermion creation and annihilation operators. Using the anticommutator properties of these operators, namely
  \begin{equation}
    \{\create_i,\create_j\}=\{\destroy_i,\destroy_i\}=0; \{\create_i,\destroy_j\}=\delta_{ij},
  \end{equation}
  and the fact that
  \begin{equation}
    \create_n\destroy_n\ket{\Phi} =
    \begin{cases}
      \ket{\Phi}, & \ket{n}\in\ket{\Phi};\\
      \ket{0}, & \ket{n}\not\in\ket{\Phi},
    \end{cases}
  \end{equation}
  we get the following possible combinations for the state $\ket*{\Phi_{ijk}^{abc}}$:
  \begin{equation}
    \ket*{\Phi_{ijk}^{abc}}=
    \begin{cases}
      \ket*{\Phi_{ijb}^{abj}}&=\ket*{\Phi_i^a},\\
      \ket*{\Phi_{ija}^{abi}}&=\ket*{\Phi_j^b},\\
      \ket*{\Phi_{ija}^{abj}}&=-\ket*{\Phi_i^b},\\
      \ket*{\Phi_{ijb}^{abi}}&=-\ket*{\Phi_j^a},\\
      \sum_{\nu}\ket*{\Phi_{ija}^{ab\nu}}&=-\sum_{\nu}\ket*{\Phi_{ij}^{b\nu}},\\
      \sum_{\nu}\ket*{\Phi_{ijb}^{ab\nu}}&=\sum_{\nu}\ket*{\Phi_{ij}^{a\nu}},\\
      \sum_{\mu}\ket*{\Phi_{ijk}^{ab\mu}}&=\sum_{\mu}\ket*{\Phi_{j\mu}^{ab}},\\
      \sum_{\mu}\ket*{\Phi_{ijk}^{ab\mu}}&=-\sum_{\mu}\ket*{\Phi_{i\mu}^{ab}}.
    \end{cases}
  \end{equation}

  We are now able to compute the matrix element \\$\vmel{\Phi_0}{\Phi_{\delta u,n}^{(1)}}$ in Eq.~\eqref{eq:bigsum} (relabelling $\mu$ as $k$ and $\nu$ as $c$),
  \begin{align}
    &\vmel{\Phi_0}{\Phi_{\delta u,n}^{(1)}}
    =\frac{\mel*{j}{\delta u}{b}}{\Delta_{bj}}\vmel{\Phi_0}{\Phi_{i}^{a}}+\frac{\mel*{i}{\delta u}{a}}{\Delta_{ai}}\vmel{\Phi_0}{\Phi_{j}^{b}}\nonumber\\
    &\hspace{4.6em}-\frac{\mel*{j}{\delta u}{a}}{\Delta_{aj}}\vmel{\Phi_0}{\Phi_{i}^{b}}-\frac{\mel*{i}{\delta u}{b}}{\Delta_{bi}}\vmel{\Phi_0}{\Phi_{j}^{a}}\nonumber\\
    &+\sum_{c\neq(a,b)}^{unocc}\Bigg\{\frac{\mel*{c}{\delta u}{b}}{\Delta_{bc}}\vmel{\Phi_0}{\Phi_{ij}^{ac}}-\frac{\mel*{c}{\delta u}{a}}{\Delta_{ac}}\vmel{\Phi_0}{\Phi_{ij}^{bc}}\Bigg\}\nonumber\\
    &+\sum_{k\neq(i,j)}^{occ}\Bigg\{\frac{\mel*{i}{\delta u}{k}}{\Delta_{ki}}\vmel{\Phi_0}{\Phi_{jk}^{ab}}-\frac{\mel*{j}{\delta u}{k}}{\Delta_{kj}}\vmel{\Phi_0}{\Phi_{ik}^{ab}}\Bigg\}\label{eq:mat1},
  \end{align}
  where $\Delta_{\alpha\beta}=\epsilon_\alpha-\epsilon_\beta$. The matrix element $\vmel{\Phi_{\delta u,0}^{(1)}}{\Phi_{n}}$ is determined in a similar manner and is given by
  \begin{align}
    &\vmel{\Phi_{\delta u,0}^{(1)}}{\Phi_{n}}
    =\frac{\mel*{j}{\delta u}{b}}{\Delta_{jb}}\vmel{\Phi_j^b}{\Phi_{ij}^{ab}}+\frac{\mel*{i}{\delta u}{a}}{\Delta_{ia}}\vmel{\Phi_i^a}{\Phi_{ij}^{ab}}\nonumber\\
    &\hspace{4.6em}-\frac{\mel*{j}{\delta u}{a}}{\Delta_{ja}}\vmel{\Phi_j^a}{\Phi_{ji}^{ab}}-\frac{\mel*{i}{\delta u}{b}}{\Delta_{ib}}\vmel{\Phi_i^b}{\Phi_{ij}^{ba}}\nonumber\\
    &+\sum_{c\neq(a,b)}^{unocc}\Bigg\{\frac{\mel*{i}{\delta u}{c}}{\Delta_{ic}}\vmel{\Phi_i^c}{\Phi_{ij}^{ab}}-\frac{\mel*{j}{\delta u}{c}}{\Delta_{jc}}\vmel{\Phi_j^c}{\Phi_{ji}^{ab}}\Bigg\}\nonumber\\
    &+\sum_{k\neq(i,j)}^{occ}\Bigg\{\frac{\mel*{k}{\delta u}{a}}{\Delta_{ka}}\vmel{\Phi_k^a}{\Phi_{ij}^{ab}}-\frac{\mel*{k}{\delta u}{b}}{\Delta_{kb}}\vmel{\Phi_k^b}{\Phi_{ij}^{ba}}\Bigg\}\label{eq:mat2}.
  \end{align}

  Finally, we compute the perturbed energy levels $E^{(1)}_{\delta u,n}$ and $E^{(1)}_{\delta u,0}$ and hence the difference $E^{(1)}_{\delta u,n}-E^{(1)}_{\delta u,0}$,
  \begin{multline} \label{eq:energies1}
    E^{(1)}_{\delta u,n}-E^{(1)}_{\delta u,0}=\mel*{\Phi_{ij}^{ab}}{\delta U}{\Phi_{ij}^{ab}}-\mel*{\Phi_0}{\delta U}{\Phi_0} = \\
    \int\dd{\vec{r}}\delta u\rvec{r} \left(|\phi_a\rvec{r}|^2+|\phi_b\rvec{r}|^2-|\phi_i\rvec{r}|^2-|\phi_j\rvec{r}|^2\right).
  \end{multline}
  We collate these terms to determine the r.h.s.\ of Eq.~\eqref{eq:bigsum}. Let us first consider what happens to the first four terms in each of Eqs.~\eqref{eq:mat1} and~\eqref{eq:mat2} in the context of Eq.~\eqref{eq:bigsum}. The contribution from the very first term in each expression is given by
  \begin{equation}
    \sum_{\substack{i,j\\i\neq j}}^{occ}\sum_{\substack{a,b\\a\neq b}}^{unocc}\vmel{\Phi_{ij}^{ab}}{\Phi_0}\frac{\mel*{j}{\delta u}{b}}{\Delta_{jb}}\bigg[\frac{\vmel{\Phi_0}{\Phi_i^a}-\vmel{\Phi_j^b}{\Phi_{ij}^{ab}}}{\Delta_{ai}+\Delta_{bj}}\Bigg],\label{eq:sglex1}
  \end{equation}
  where
  \begin{align}
    \vmel{\Phi_0}{\Phi_i^a}-\vmel{\Phi_j^b}{\Phi_{ij}^{ab}}&=\Big(\sum_{k\in\Phi_u}-\sum_{k\in\Phi_i^a}\Big)\mell{ik}{ak}\\
    &=\mell{ij}{aj}-\mell{ib}{ab},
  \end{align}
  with $\mell{ij}{ab}=\mel*{ij}{V_{ee}}{ab}-\mel*{ij}{V_{ee}}{ba}$. The other three terms in Eqs.~\eqref{eq:mat1} and~\eqref{eq:mat2} which involve a single-orbital substitution can be evaluated in a similar manner, and by relabelling dummy indices it can be shown that each of terms is equal. The total contribution from these terms is therefore
  \begin{equation}
    4\sum_{\substack{i,j\\i\neq j}}^{occ}\sum_{\substack{a,b\\a\neq b}}^{unocc}\vmel{\Phi_{ij}^{ab}}{\Phi_0}\frac{\mel*{i}{\delta u}{a}}{\Delta_{ai}}\frac{\mell{ji}{bi}-\mell{ja}{ba}}{\Delta_{ai}+\Delta_{bj}}.
  \end{equation}
  It is noted that several of the other terms in Eqs.~\eqref{eq:mat1} and~\eqref{eq:mat2} are duplicates of each other, which again can be seen by relabelling dummy indices. After expanding all the remaining terms in Eqs.~\eqref{eq:mat1},~\eqref{eq:mat2} and~\eqref{eq:sglex1} 
  in terms of KS orbitals, the functional derivative of the double excitations term is found to be equal to
  \begin{align}
    &\fdv{D[u]}{u\r}=2\sum_{\substack{i,j\\i\neq j}}^{occ}\sum_{\substack{a,b\\a\neq b}}^{unocc}\frac{\mell{ab}{ij}}{\Delta_{ai}+\Delta_{bj}}\nonumber \\
    &\Bigg\{2\phi_i^*\r\phi_a\r\frac{\mell{ji}{bi}-\mell{ja}{ba}}{\Delta_{ai}}\nonumber\\
    &+\sum_{c\neq(a,b)}^{unocc}\phi_c^*\r\phi_b\r\frac{\mell{ij}{ac}}{\Delta_{bc}}+\phi_i^*\r\phi_c\r\frac{\mell{cj}{ab}}{\Delta_{ic}}\nonumber\\
    &+\sum_{k\neq(i,j)}^{occ}\phi_i^*\r\phi_k\r\frac{\mell{jk}{ab}}{\Delta_{ki}}-\phi_k^*\r\phi_a\r\frac{\mell{ij}{kb}}{\Delta_{ka}}\nonumber\\
    &-\frac{1}{2}\big[|\phi_a\r|^2-|\phi_i\r|^2\big]\frac{\mell{ij}{ab}}{\Delta_{ai}+\Delta_{bj}}\Bigg\}+\cc \label{eq:matorbitals1}
  \end{align}
  The above expression is equal to zero at the minimizing potential, $u\r=u_0\r$.

    This result is reminiscent of the derivative of the double-excitations part of the second-order 
    correlation energy in traditional DFT PT. 
    In Ref.~\cite{Sanchez_Wu}, in which part of the KS potential is expanded in terms of a 
    Gaussian basis set $\{g_t\r\}$ with coefficients $b_t^\sigma$, the derivative of the doubly-excited correlation energy 
    term with respect to $b_t^\sigma$ can be expressed as
    \begin{equation}
      \pdv{E_d^{(2)}}{b_t^\sigma}=-\int\dd{\vec{r}}g_t\r \fdv{D[u]}{u\r},
    \end{equation}
    with 
    $\delta D [ u ] / \delta u \r $ given by Eq.~\eqref{eq:matorbitals1}.
    However, as previously stressed, in Ref.~\cite{Sanchez_Wu} 
    and other 
    works in DFT PT, the minimization is carried out over the total energy, which is unbound from below. 
    We discuss at some length the issues with a total energy minimization 
    using a second-order correlation energy functional in section \ref{sec:5.4}.

  We can further simplify Eq.~\eqref{eq:matorbitals1} in a manner which is also beneficial if we want to employ the Uns\"old approximation~\cite{unsold1927} (common energy denominator approximation)~\cite{unsold1927,kli1,kli2,CEDA,localizedHF,elp_2006,elp_2007}.
  We note that some terms contain a denominator of mixed sign, which
  yields less accurate results if we approximate the denominators with a constant.
  Consider the complex conjugate of the expression
  \begin{align}
    &\Bigg[\sum_{\substack{i,j\\i\neq j}}^{occ}\sum_{\substack{a,b,c\\a\neq b\neq c}}^{unocc}\phi_c^*\r\phi_b\r\frac{\mell{ab}{ij}}{\Delta_{ai}+\Delta_{bj}}\frac{\mell{ij}{ac}}{\Delta_{bc}}\Bigg]^*\nonumber\\
    &=\sum_{\substack{i,j\\i\neq j}}^{occ}\sum_{\substack{a,b,c\\a\neq b\neq c}}^{unocc}\phi_c\r\phi^*_b\r\frac{\mell{ij}{ab}}{\Delta_{ai}+\Delta_{bj}}\frac{\mell{ac}{ij}}{\Delta_{bc}}\nonumber\\
    &=\sum_{\substack{i,j\\i\neq j}}^{occ}\sum_{\substack{a,b,c\\a\neq b\neq c}}^{unocc}\phi_c^*\r\phi_b\r\frac{\mell{ab}{ij}}{\Delta_{ai}+\Delta_{cj}}\frac{\mell{ij}{ac}}{\Delta_{cb}},
  \end{align}
  where in the last step we have just swapped the labels of the dummy indices $b$ and $c$. This term plus its complex conjugate is therefore equal to
  \begin{align}
    \sum_{\substack{i,j\\i\neq j}}^{occ}\sum_{\substack{a,b,c\\a\neq b\neq c}}^{unocc}&\phi_c^*\r\phi_b\r\mell{ab}{ij}\frac{\mell{ij}{ac}}{\Delta_{bc}}\times\nonumber\\
    &\Bigg[\frac{1}{\Delta_{ai}+\Delta_{bj}}-\frac{1}{\Delta_{ai}+\Delta_{cj}}\Bigg]\nonumber\\
    =-\sum_{i,j}^{occ}\sum_{a,b,c}^{unocc}&\phi_c^*\r\phi_b\r\frac{\mell{ab}{ij}\mell{ij}{ac}}{(\Delta_{ai}+\Delta_{bj})(\Delta_{ai}+\Delta_{cj})}, \label{eq:mixed1}
  \end{align}
  where the denominator is now of fixed (positive) sign. We can perform a similar procedure for the term with denominator $\Delta_{ki}$, which with its complex conjugate becomes
  \begin{equation}
    -\sum_{i,j,k}^{occ}\sum_{a,b}^{unocc}\phi_i^*\r\phi_k\r\frac{\mell{ab}{ij}\mell{jk}{ab}}{(\Delta_{ai}+\Delta_{bj})(\Delta_{ak}+\Delta_{kj})}. \label{eq:mixed2}
  \end{equation}
  Using Eqs.~\eqref{eq:mixed1} and~\eqref{eq:mixed2}, we can rewrite Eq.~\eqref{eq:matorbitals1} as
  \begin{align} 
    &\fdv{D[u]}{u\r}=\sum_{i,j}^{occ}\sum_{a,b}^{unocc}\frac{\mell{ab}{ij}}{\Delta_{ai}+\Delta_{bj}}\nonumber\\
    &\Bigg\{4\phi_i^*\r\phi_a\r\frac{\mell{ji}{bi}-\mell{ja}{ba}}{\Delta_{ai}}\nonumber\\
    &-\sum_{c\neq(a,b)}^{unocc}\Bigg[\phi_c^*\r\phi_b\r\frac{\mell{ij}{ac}}{\Delta_{ai}+\Delta_{cj}}-2\phi_i^*\r\phi_c\r\frac{\mell{cj}{ab}}{\Delta_{ci}}\Bigg]\nonumber\\
    &-\sum_{k\neq(i,j)}^{occ}\Bigg[\phi_i^*\r\phi_k\r\frac{\mell{jk}{ab}}{\Delta_{ak}+\Delta_{bj}}+2\phi_k^*\r\phi_a\r\frac{\mell{ij}{kb}}{\Delta_{ak}}\Bigg]\nonumber\\
    &-\big[|\phi_a\r|^2-|\phi_i\r|^2\big]\frac{\mell{ij}{ab}}{\Delta_{ai}+\Delta_{bj}}\Bigg\}+\cc
    \label{63}
  \end{align}
  If desired, it is now straightforward to use the Uns\"old approximation~\cite{unsold1927} 
  (set all denominators $\Delta$ equal to a constant) and remove the summations over the unoccupied orbitals using the 
  completeness relation.

  The minimizing potential $u_0$ is determined by setting the functional derivative (\ref{eq:matorbitals1},~\ref{63}) to zero:
  \begin{equation} \label{eq66}
    \left. \fdv{D[u]}{u\r} \right|_{u = u_0}= 0 \, .
  \end{equation}
   Eq.~\eqref{eq66} must be solved iteratively with an
  energy minimization algorithm such as steepest
  descent. At the $n^\text{th}$ iteration, the
  potential will be $u^{(n)} \r$. Substituting the single-particle
  orbitals $\phi_{u,p}^{(n)} \r$ and energies
$\epsilon_{u,p}^{(n)}$ of $u^{(n)} \r$ into~\eqref{63}, we
obtain 
$ \delta D [u] / \delta u \r $ at $u^{(n)}$. Using this functional
derivative we correct the potential, $u^{(n)} \rightarrow $
$u^{(n+1)}$, so as to lower $D[u]$.  Finally, we iterate until the
functional derivative~\eqref{eq66} vanishes.

  Once the optimal potential $u_0$ has been found, together with its single-particle orbitals $\phi_{u_0, p}$ and 
  energies $\epsilon_{u_0 , p}$,  
  we may proceed to determine the first-order KS potential by minimizing $S_{u_0} [w]$ over $w$, keeping $u_0$ fixed.
  
  The minimizing potential $w_0 [u_0] = u_0 + v' [u_0]$~\eqref{eq12} is given by (for fixed $u_0$):
  \begin{multline}
    \label{eq67}
    0 = \left. {\delta S_{u_0}[ w ]  \over \delta w ({\bf r}) } \right|_{ w=u_0 + v' [u_0]}
    = \\
    \sum_{i, \, a }
        { \langle \phi_{{u_0},i} | {\cal J}_{u_0}  - {\cal K}_{u_0} -
          u_0 - v' [ u_0 ] | \phi_{{u_0},a} \rangle
          \over {\epsilon_{{u_0} , i} - \epsilon_{{u_0} , a}} } \times \\
        \phi_{{u_0} , a}^*  ({\bf r})  \phi_{{u_0} , i} ({\bf r})
        + {\rm c.c.}
  \end{multline}
  Eq.~\eqref{eq67} is a standard OEP equation for the potential $v'[u_0]$ with the simplification that during the 
  solution of the OEP equation the orbitals $\phi_{u_0, p}$ and their energies $\epsilon_{u_0 , p}$ remain fixed and 
  independent of $v' [ u_0]$.
    
  The first-order correction $v' [ u_0] $ does not vanish.
  %
 Finally, the KS potential to first order is given by
\begin{equation} \label{eq68} 
  v_s [ u_0 ] \r = v_{en} \r + u_0 \r + \alpha
  v' [ u_0 ] \r + {\cal O}(\alpha^2) ;
\end{equation}
the correlation energy corresponding to the KS potential is
\begin{equation}
  E_{u_0}^c \big[ u_0 + v' [ u_0 ] \big] = - S_{u_0} \big[ u_0 + v' [u_0] \big] - D [u_0] .
\end{equation}

The criterion for the validity of the approximation in \eqref{20} and  \eqref{21}, in which $S_{u} \big[ u + v' [u] \big]$ is neglected in
the minimization of $T_u \big[ w_0 [u] \big]$, is
\begin{equation} \label{eq69} 
S_{u_0} \big[ u_0 + v' [u_0] \big] \ll D [u_0] .
\end{equation}

  In summary, by minimizing $T_u [w]$ over $u$ and $w$, not only is the magnitude of the correlation energy the smallest possible over all
  $u$ and $w$, leading to a fast converging expansion of the KS potential,
  but also the resulting first order KS potential $v_s [u_0]$ has both exchange and correlation character, rather than just exchange.

  \subsection{Analysis of total energy minimization in DFT PT using a second-order correlation functional} \label{sec:5.4}
  
In this section, we focus on the functional derivative (f.d.) of the second-order correlation energy functional and of the total energy in DFT PT 
and analyse the tendency to variational collapse that has been observed in calculations.
 
Using notation in this paper, \eqref{eq8} and  \eqref{17b}, the second-order correlation energy, $E_c [\rho]$, in DFT PT 
\cite{GL_PT_1,GL_PT_2} is
given by
  \begin{equation} \label{eq71}
E_c [\rho]  =  E^c_{v_{Hxc [\rho] }} \big[ v_{Hx} [\rho] \big]  ,
  \end{equation}
where $v_{Hxc} [\rho]$ is the Hartree, exchange and correlation part of the KS potential with density $\rho$. 
We note that just the Hx part of the KS potential of density $\rho$ 
appears in the argument of the correlation energy functional on the right 
(in the square brackets, amounting to $w = v_{Hx}$ in \eqref{eq8}), 
although the KS orbitals and their energies are obtained from the KS equations with the full Hxc potential 
(which gives the dependence in the subscript, i.e., $u = v_{Hxc}$ in \eqref{eq8}).   
Some authors use the simpler form, where both potentials are the same~\cite{Sanchez_Wu}:
  \begin{equation} \label{eq72}
E_c [\rho]  =  E^c_{v_{Hxc [\rho] }} \big[ v_{Hxc} [\rho] \big]  .
  \end{equation}
To proceed with the analysis and compare with our method, it is convenient to view the density functionals 
\eqref{eq71} and \eqref{eq72} as potential functionals. 
Hence, we consider the density, $\rho = \rho_u$, to be the g.s. density of an effective Hamiltonian $H_u$, 
with g.s. Slater determinant $\Phi_u$, see Eqs.~\eqref{eq2}-\eqref{3}.
The effective potential $u$ is the Hxc potential and from \eqref{eq18} the Hx part of the KS potential with density $\rho_u$ 
is $w_0 [ u ]$~\footnote{$w_0 [ u ]$ is the Hx part of the KS potential with density $\rho_u$, and emulates the Hxc part of the KS 
potential with density $\rho_s [u]$ \eqref{eq16}. See section \ref{sec:4.2}.}. 
Finally, the second-order correlation energy of DFT PT \eqref{eq71} can be written as a potential functional~\cite{yang_2004}, 
using our notation, as  
\begin{equation}
E_c [\rho_u]  =  E^c_{u} \big[ w_0 [u] \big] = - S_{ u } [ w_0 [ u ] \big] - D [ u ] ; \label{eq73} 
\end{equation}
and DFT's total energy (as a potential-functional) is
\begin{equation} \label{eq74}
E [\rho_u] = \langle \Phi_u | H | \Phi_u \rangle - S_{ u } \big[ w_0[ u ] \big] - D [ u ] .
\end{equation}
Using the simpler form for the correlation energy \eqref{eq72}, we have
\begin{equation} \label{eq75}
E_c [\rho_u]  =  E^c_{ u } [ u ]  = - S_{ u } [ u ] - D [ u ] ;
\end{equation}
and the corresponding total energy potential-functional is
\begin{equation} \label{eq76}
E [\rho_u] = \langle \Phi_u | H | \Phi_u \rangle - S_{ u } [ u ] - D [ u ].
\end{equation}
It is common practice with potential functionals (or equivalently implicit density functionals) to employ the OEP method 
to minimize the total energy. 
The functional derivative of the total energy w.r.t. the effective potential is, in the two cases: 
\begin{equation} \label{eq77}
{\delta E [\rho_u]  \over \delta u \r} = 
{\delta \over \delta u \r} \langle \Phi_u | H | \Phi_u \rangle  
- {\delta S_{ u } \big[ w_0[ u ] \big] \over \delta u \r} - {\delta D [ u ] \over \delta u \r} 
\end{equation}
and
\begin{equation} \label{eq78}
{\delta E [\rho_u]  \over \delta u \r} = 
{\delta \over \delta u \r} \langle \Phi_u | H | \Phi_u \rangle  
- {\delta S_{ u } [ u  ] \over \delta u \r} - {\delta D [ u ] \over \delta u \r} .
\end{equation}
To simplify the two functional derivatives, first we note the identity~\cite{gidopoulos1}:
\begin{equation} \label{eq79}
{\delta \over \delta u \r} \langle \Phi_u | H | \Phi_u \rangle = \left. {\delta S_u[w] \over \delta w ({\bf r}) } \right|_{ u} .
\end{equation}
Using the chain rule we have
\begin{multline} \label{eq80}
 {\delta S_{ u } \big[ w_0[ u ] \big] \over \delta u \r}  = 
\left. {\delta S_{ u } [ w ]  \over \delta u \r} \right|_{w_0[ u ]}  + \\
 \int d {\bf x} \,  {\delta w_0 [u] ({\bf x}) \over \delta u \r} \, 
\left. {\delta S_{ u } [ w ] \over \delta w ({\bf x} )} \right|_{w_0[u]} .
\end{multline}
From \eqref{eq80}, \eqref{eq18} and \eqref{17}, we obtain
\begin{equation} \label{eq81}
{\delta S_{ u } \big[ w_0[ u ] \big] \over \delta u \r} = 
\left. {\delta S_{ u } [ w ]  \over \delta u \r} \right|_{w_0[ u ]}.
\end{equation}
We conclude that the f.d. of the total energy \eqref{eq74} is
\begin{equation} \label{eq82}
{\delta E [\rho_u]  \over \delta u \r} = 
\left. {\delta S_u[w] \over \delta w ({\bf r}) } \right|_{  u} -
\left. {\delta S_{ u } [ w ]  \over \delta u \r} \right|_{w_0[ u ]} - {\delta D [ u ] \over \delta u \r} ;
\end{equation}
and the f.d. of the total energy \eqref{eq76} is
\begin{equation} \label{eq83}
{\delta E [\rho_u]  \over \delta u \r} = 
 - \left. {\delta S_{ u } [ w ]  \over \delta u \r} \right|_{u} - {\delta D [ u ] \over \delta u \r} .
\end{equation}
The f.d. of the total energy \eqref{eq74} is the sum of three terms \eqref{eq82}. The first term vanishes for the xOEP potential $u_{Hx}$, see section \ref{sec:5.1}. 
The sum of the second and third terms vanishes for $u_0$, the minimizing potential of $T_u \big[ w_0[u] \big]$. 
Hence, the total energy \eqref{eq74} will have a stationary point (but not a minimum) at a potential lying somewhere between $u_{Hx}$ and $u_0$. That potential will be the Hxc potential of DFT PT. 
It is intriguing to investigate the relation of the latter potential with the Hxc potential of the present theory, 
$u_0 + \alpha v ' [ u_0 ]$ \eqref{eq68}.  

The minimization of the total energy \eqref{eq74} over $u$ amounts to a balanced search to achieve two goals: to minimize the expectation value $\langle \Phi_u | H | \Phi_u \rangle$ (well behaved) and to maximize the second-order difference  
$T_u \big[ w_0[u] \big]$. 
Although bound from below, $T_u \big[ w_0[u] \big]$ is not bound from above and the search will be biased towards the maximization of $T_u \big[ w_0[u] \big]$. During the iterations the potential is expected to move away from the minimum of $T_u \big[ w_0[u] \big]$. Hence, the second term on the r.h.s. of \eqref{eq82}, which we had omitted based on \eqref{20}, \eqref{21}, can no longer be neglected as it is prone to diverge, similarly to the third term. 

The f.d. of the total energy \eqref{eq75} has only two terms \eqref{eq83} because the f.d. of 
$\langle \Phi_u | H | \Phi_u \rangle$ cancels with part of the f.d. of $S_u[u]$ \eqref{eq79}. 
Thus, fully self-consistently and without risk of variational collapse, 
the Hxc potential (solution of $\delta E[\rho_u] /$ $ \delta u \r = 0$) 
can be obtained by searching for the potential $\tilde u $ (dependent on $w$) that minimizes the (positive) 
second-order quantity $S_u[w] + D[u] $ and then choosing $w$ so that $\tilde u = w$. 
From \eqref{eq76} and \eqref{eq83}, it is evident that an algorithm to minimize $S_u[w] + D[u]$ 
will effectively maximize rather than minimize the total energy \eqref{eq76}.
Even more strongly than the previous case, the minimization of the total energy \eqref{eq75} does little to lower the value of $\langle \Phi_u | H | \Phi_u \rangle$ (since the f.d. of this term cancels) while it leads to the divergence of $S_u[w] + D[u]$.

  \section{Summary and Discussion}


  The research reported in this paper builds on previous work at the interface between wave function theory (WFT) and Kohn-Sham
  (KS) density functional theory (DFT)~\cite{gidopoulos1}.
  %
  The link between WFT and KS-DFT, established in~\cite{gidopoulos1}, is that among all non-interacting Hamiltonians $H_v$
  with an effective potential $v \r$, {\em the KS effective Hamiltonian adopts energetically optimally the interacting ground state as
    its approximate ground state.}
  %
  Specifically, the KS potential turns out to be optimal in that it minimizes an appropriate energy difference $T_\Psi [ v ] $~\eqref{eq1}
  over all effective potentials $v \r$.
  %
  This energy difference depends on the interacting state $\Psi$ and is strictly positive, $T_\Psi [ v ] > 0$~\eqref{eq1}.

  There is a large number of partially interacting Hamiltonians $H_u (\alpha)$~\eqref{4}, with $0 \le \alpha \le 1$,
  that yield the interacting Hamiltonian of interest $H$ for $\alpha = 1$;
  they differ in the choice of effective potential $u \r$ appearing in the zero-order Hamiltonian $H_u$~\eqref{eq2}.
  For any of these partially/weakly interacting systems of electrons, their ground state $\Psi_u (\alpha)$ can be expanded
  in a power series in the small perturbation
  $\alpha \big( V_{ee} - \sum_i u ({\bf r}_i) \big) $.
  When we replace $\Psi$ in the energy difference $T_\Psi [ v ] $, with an expansion of any of the partially interacting ground states
  $\Psi_u (\alpha)$, truncated at a finite order, we obtain a corresponding power series expansion of the energy difference.
  Minimizing order-by-order the expansion of the energy difference w.r.t.\ the effective potential $v$,
  we obtain a corresponding power series expansion of the KS potential in powers of $\alpha$.

  There are at least as many expansions of the KS potential in powers of $\alpha$ as there are choices for the zero-order potential $u$.
  For any of the weakly interacting ground states $\Psi_u ( \alpha ) $, and for small $\alpha$, the dominant term in the expansion of the energy
  difference $T_\Psi [ v ] $ is second order: $ T_{\Psi_u (\alpha)} [ u + \alpha v ' ] = \alpha^2 \, T_u [ w ] + {\cal O} (\alpha^3) $, with
  $ w = u + v'$ and $T_u [ w ] > 0$.

  Minimization of the second order energy difference $T_u [ w ] $ over $w \r $ gives an expansion of the KS potential up to first order.
  The aim is to choose optimally the zero-order effective potential $u$ in order to obtain fast converging expansions for
  $\Psi_u ( \alpha) $ and for the KS potential.

  The link between WFT and KS-DFT is explored further in the present work:
  We consider the correlation energy $E_H^c [ v ] $ of the interacting system, with non-interacting reference the ground state
  $\Phi_v $ of the Hamiltonian with effective potential $v $.
  The potential that minimizes the magnitude of the correlation energy $E_H^c [ v ] $ over all effective potentials $v$ is xOEP.

  When we expand the ground state energy of the partially interacting system in powers of $\alpha$, we obtain a power series expansion of
  the correlation energy.
  We consider the correlation energy $E_{H_u ( \alpha) }^c [ u + \alpha v' ] $ of the partially interacting system with reference to the ground
  state of the effective potential $u \r + \alpha v' \r$. In the weakly interacting limit, $\alpha \rightarrow 0$, the dominant term in the
  expansion of the correlation energy is second order and it is equal to minus the second-order energy difference:
  $E_{H_u ( \alpha) }^c [ u + \alpha v' ]  = - \alpha^2 \, T_u [ w ] + {\cal O} (\alpha^3)$, with $w = u + v'$, see Eq.~\ref{eq22}.
  This is the first important result of the paper.

  We recall that the optimization of the energy difference $T_u [ w ]$ over all effective potentials $w$ (i.e.\ over all reference ground states
  of $w$) yields the KS potential up to first-order.
  We conclude that, for any $u$, the ground state of the KS potential (up to first-order)
  is the optimum reference for the correlation energy,
  since the magnitude of the second order correlation energy is minimum for that reference.
  We extend this reasoning by seeking the effective potential $u$ for which the correlation energy from the KS reference state
  $E_u^c \big[ w_0 [u] \big] = - T_u \big[ w_0 [u] \big]$ (already a quantity with minimum magnitude over $w$)
  also has small or minimum magnitude over the zero-order potential $u$.

  Intuitively, small magnitude of correlation energy implies weak perturbation and hence fast convergence of the perturbative expansion for
  $\Psi_u (\alpha)$ and for the KS potential.

  We consider three choices for the zeroth-order potential $u$.
  In the first two, the density of the zero order state is equal to the density of the weakly interacting state, within first order.
  In both cases, the first order term in the expansion of the KS potential vanishes.
  These two choices yield the Hartree and exact exchange potential of DFT (xOEP) and the Hartree and LFX potential~\cite{LFX_Hollins}.

  By minimizing the magnitude of the correlation energy over $w$ and over $u$ (our third choice) we hope to obtain the fastest converging
  power series expansion for $\Psi_u (\alpha)$ and for the KS potential, with the latter having exchange and correlation character.
  Since our second order expressions are bound from below, their minimization is mathematically well posed. We claim then that we have derived
  for the first time well behaved equations determining in an ab initio manner the KS potential with Hartree,
  exchange and correlation character, in a power series expansion of the potential up to first order.


  \section{Acknowledgement}
  NIG acknowledges financial support by The Leverhulme Trust, through a Research
  Project Grant with number RPG-2016-005.

  \section{Authors contributions}
  
  Both authors contributed to the research and to the writing of this paper.
 
  %
  \bibliographystyle{epj}
  \bibliography{bibliography}

\begin{thebibliography}{52}

\bibitem{Drugs}
D.~Li, Y.~Wang, K.~Han, Coord. Chem. Rev. \textbf{256}, 1137 (2012)

\bibitem{superconductivity}
D.~Duan, Y.~Liu, F.~Tian, D.~Li, X.~Huang, Z.~Zhao, H.~Yu, B.~Liu, W.~Tian,
  T.~Cui, Sci. Rep. \textbf{4} (2014)

\bibitem{Burke_review}
K.~Burke, J. Chem. Phys. \textbf{136}, 150901 (2012)

\bibitem{Kohn_Nobel_lecture}
W.~Kohn, Rev. Mod. Phys. \textbf{71}, 1253 (1999)

\bibitem{Pople_Nobel_lecture}
J.A. Pople, Rev. Mod. Phys. \textbf{71}, 1267 (1999)

\bibitem{engel1999explicit}
E.~Engel, R.M. Dreizler, J. Comp. Chem. \textbf{20}, 31 (1999)

\bibitem{Engel_vdw}
E.~Engel, A.~H\"ock, R.M. Dreizler, Phys. Rev. A \textbf{61}, 032502 (2000)

\bibitem{Grabowski2}
I.~Grabowski, S.~Hirata, S.~Ivanov, R.J. Bartlett, J. Chem. Phys \textbf{116},
  4415 (2002)

\bibitem{abinitioDFT}
R.J. Bartlett, V.F. Lotrich, I.V. Schweigert, J. Chem. Phys. \textbf{123},
  062205 (2005)

\bibitem{Sanchez_Wu}
P.~Mori-S\`anchez, Q.~Wu, W.~Yang, J. Chem. Phys. \textbf{123}, 062204 (2005)

\bibitem{Schweigert_ab_init_correlation_funcs}
I.V. Schweigert, V.F. Lotrich, R.J. Bartlett, J. Chem. Phys. \textbf{125},
  104108 (2006)

\bibitem{grabowski}
I.~Grabowski, V.~Lotrich, R.J. Bartlett, J. Chem. Phys \textbf{127}, 154111
  (2007)

\bibitem{Bartlett}
R.J. Bartlett, Mol. Phys. \textbf{108}, 3299 (2010)

\bibitem{Levy_constrained}
M.~Levy, Proc. Natl. Acad. Sci. USA \textbf{76}, 6062 (1979)

\bibitem{Lieb_constrained}
E.H. Lieb, Int. J. of Quant. Chem. \textbf{24}, 243 (1983)

\bibitem{ad_conn_1}
J.~Harris, R.O. Jones, J. Phys. F \textbf{4}, 1170 (1974)

\bibitem{ad_conn_2}
D.~Langreth, J.~Perdew, J. Solid State Commun. \textbf{17}, 1425  (1975)

\bibitem{ad_conn_3}
O.~Gunnarsson, B.I. Lundqvist, Phys. Rev. B \textbf{13}, 4274 (1976)

\bibitem{GL_PT_1}
A.~G\"orling, M.~Levy, Phys. Rev. B \textbf{47}, 13105 (1993)

\bibitem{GL_PT_2}
A.~G\"orling, M.~Levy, Phys. Rev. A \textbf{50}, 196 (1994)

\bibitem{OEP1}
R.T. Sharp, G.K. Horton, Phys. Rev. \textbf{90}, 317 (1953)

\bibitem{OEP2}
J.D. Talman, W.F. Shadwick, Phys. Rev. A \textbf{14}, 36 (1976)

\bibitem{rohr2006variational}
D.~Rohr, O.~Gritsenko, E.~Baerends, Chem. Phys. Lett. \textbf{432}, 336 (2006)

\bibitem{bartlett_collapse}
D.~Bokhan, R.J. Bartlett, Chem. Phys. Lett. \textbf{427}, 466 (2006)

\bibitem{LANGRETH19751425}
D.~Langreth, J.~Perdew, Solid State Commun. \textbf{17}, 1425  (1975)

\bibitem{Perdew_RPA}
S.~Kurth, J.P. Perdew, Phys. Rev. B \textbf{59}, 10461 (1999)

\bibitem{Bartlett_RPA}
P.~Verma, R.J. Bartlett, J. Chem. Phys. \textbf{136}, 044105 (2012)

\bibitem{Gross_RPA}
M.~Hellgren, D.R. Rohr, E.K.U. Gross, J. Chem. Phys. \textbf{136}, 034106
  (2012)

\bibitem{Gorling_RPA_2012}
P.~Bleiziffer, A.~He{\ss}elmann, A.~Görling, J. Chem. Phys. \textbf{136}, 134102
  (2012)

\bibitem{Gorling_RPA2}
P.~Bleiziffer, M.~Krug, A.~Görling, J. Chem. Phys. \textbf{142}, 244108 (2015)

\bibitem{nguyen_2009}
H.V. Nguyen, S.~de~Gironcoli, Physical Review B \textbf{79}, 205114 (2009)

\bibitem{RPA_review}
G.P. Chen, V.K. Voora, M.M. Agee, S.G. Balasubramani, F.~Furche, Annual Review
  of Physical Chemistry \textbf{68}, 421 (2017), pMID: 28301757

\bibitem{rinke_2012}
X.~Ren, P.~Rinke, C.~Joas, M.~Scheffler, Journal of Materials Science
  \textbf{47}, 7447 (2012)

\bibitem{kubo}
R.~Kubo, Rep. Prog. Phys. \textbf{29}, 255 (1966)

\bibitem{godby_sham_schluter}
R.W. Godby, M.~Schl\"uter, L.J. Sham, Phys. Rev. B \textbf{37}, 10159 (1988)

\bibitem{sham_schluter}
L.~Sham, M.~Schl{\"u}ter, Phys. Rev. Lett. \textbf{51}, 1888 (1983)

\bibitem{gidopoulos1}
N.I. Gidopoulos, Phys. Rev. A \textbf{83}, 040502 (2011)

\bibitem{lieb_1983}
E.H. Lieb, International journal of quantum chemistry \textbf{24}, 243 (1983)

\bibitem{teale_2017}
T.J. Irons, J.W. Furness, M.S. Ryley, J.~Zemen, T.~Helgaker, A.M. Teale, The
  Journal of chemical physics \textbf{147}, 134107 (2017)

\bibitem{gross_book}
E.K. Gross, R.M. Dreizler, \emph{{Density Functional Theory}}, Vol. 337
  (Springer Science \& Business Media, 2013)

\bibitem{engel_dreizler}
E.~Engel, R.M. Dreizler, \emph{{Density Functional Theory: An Advanced Course}}
  (Springer Science \& Business Media, 2011)

\bibitem{LFX_Hollins}
T.~Hollins, S.~Clark, K.~Refson, N.~Gidopoulos, Journal of Physics: Condensed
  Matter \textbf{29}, 04LT01 (2016)

\bibitem{levy_pnas}
M.~Levy, Proc. Natl. Acad. Sci. USA \textbf{76}, 6062 (1979)

\bibitem{payne_1979}
P.W. Payne, J. Chem. Phys. \textbf{71}, 490 (1979)

\bibitem{staroverov2013}
I.G. Ryabinkin, A.A. Kananenka, V.N. Staroverov, Phys. Rev. Lett. \textbf{111},
  013001 (2013)

\bibitem{szabo2012modern}
A.~Szabo, N.S. Ostlund, \emph{Modern {Q}uantum {C}hemistry} (New York:
  Macmillan, 2012)

\bibitem{OEP_Hylleraas}
T.W. Hollins, S.J. Clark, K.~Refson, N.I. Gidopoulos, Phys. Rev. B \textbf{85},
  235126 (2012)

\bibitem{OEP_Kummel}
S.~K\"ummel, J.P. Perdew, Phys. Rev. Lett. \textbf{90}, 043004 (2003)

\bibitem{Gido_Lath}
N.I. Gidopoulos, N.N. Lathiotakis, Phys. Rev. A \textbf{85}, 052508 (2012)

\bibitem{unsold1927}
A.~Uns{\"o}ld, Zeitschrift f{\"u}r Physik \textbf{43}, 563 (1927)

\bibitem{CEDA}
M.~Gr{\"u}ning, O.V. Gritsenko, E.~Baerends, The Journal of Chemical Physics
  \textbf{116}, 6435 (2002)

\bibitem{yang_2004}
W.~Yang, P.W. Ayers, Q.~Wu, Physical review letters \textbf{92}, 146404 (2004)

\end{thebibliography}

  %

\end{document}